\documentclass{article}

\usepackage{fullpage}
\usepackage{tikz}
\usepackage[all]{xy}
\usepackage{color}
\usepackage{float}
\usepackage{graphicx}
\usepackage{amsmath}
\usepackage{wrapfig}
\usepackage{epstopdf}
\usepackage{amssymb}
\usepackage{authblk}
\usepackage{boxedminipage}
\usepackage{framed}

\definecolor{Sion}{rgb}{.45,0.05,.85}

\def\defn#1{\textbf{\textit{#1}}}

\title{The importance of the whole: topological data analysis for the network neuroscientist}

\author[1]{Ann E. Sizemore}
\author[1]{Jennifer E. Phillips-Cremins}
\author[2,4]{Robert Ghrist}
\author[1,3,4,5,*]{Danielle S. Bassett}

\affil[1]{Department of Bioengineering, School of Engineering and Applied Sciences, University of Pennsylvania, Philadelphia, USA}
\affil[2]{Department of Mathematics, College of Arts and Sciences, University of Pennsylvania, Philadelphia, USA}
\affil[3]{Department of Physics \& Astronomy, College of Arts and Sciences,University of Pennsylvania, Philadelphia, USA}
\affil[4]{Department of Electrical \& Systems Engineering, School of Engineering and Applied Sciences, University of Pennsylvania, Philadelphia, USA}
\affil[5]{Department of Neurology, Perelman School of Medicine, University of Pennsylvania, Philadelphia, USA}

\begin{document}
\maketitle

\begin{abstract}
The application of network techniques to the analysis of neural data has greatly improved our ability to quantify and describe these rich interacting systems. Among many important contributions, networks have proven useful in identifying sets of node pairs that are densely connected and that collectively support brain function. Yet the restriction to pairwise interactions prevents us from realizing intrinsic topological features such as cavities within the interconnection structure that may be just as crucial for proper function. To detect and quantify these topological features we must turn to methods from algebraic topology that encode data as a simplicial complex built of sets of interacting nodes called simplices. On this substrate, we can then use the relations between simplices and higher-order connectivity to expose cavities within the complex, thereby summarizing its topological nature. Here we provide an introduction to persistent homology, a fundamental method from applied topology that builds a global descriptor of system structure by chronicling the evolution of cavities as we move through a combinatorial object such as a weighted network. We detail the underlying mathematics and perform demonstrative calculations on the mouse structural connectome, electrical and chemical synapses in \textit{C. elegans}, and genomic interaction data. Finally we suggest avenues for future work and highlight new advances in mathematics that appear ready for use in revealing the architecture and function of neural systems.
\end{abstract}

\section*{Introduction}

Network science now branches far into applied fields such as medicine \cite{barabasi2011network}, genetics \cite{alon2007network}, physics \cite{papadopoulos2017network}, sociology \cite{wasserman1994advances}, ecology \cite{proulx2005network}, and neuroscience \cite{sporns2014contributions}. Such breadth is made possible by the discipline's roots in the abstract field of graph theory. Using this deep theoretical foundation, we now analyze network models with an ease and finesse previously not possible, while confidently continuing to develop the underlying mathematical framework. For the most elementary application of graph-based methods, we only require our system to be composed of units and their pairwise relations; a simple requirement that is commonly met in the modeling of many biological systems, including the brain. Yet often still another level of organization exists in which groups of units all serve a function together (perhaps via a similar process or capacity), implying that any subset of actors in a group also has this relation. Many studies spanning cellular and areal scales stress the importance of such higher-order interactions \cite{ganmor2011sparse,bassett2014cross}. Algebraic topology presents us with a language with which to encode, study, and manipulate such structures with higher-order relations. In addition, the field of algebraic topology gifts us with nearly 100 years of theoretical groundwork distinct from and yet complementary to that offered by graph theory.

\begin{figure}[h]
	\centering
	\includegraphics[width=5in]{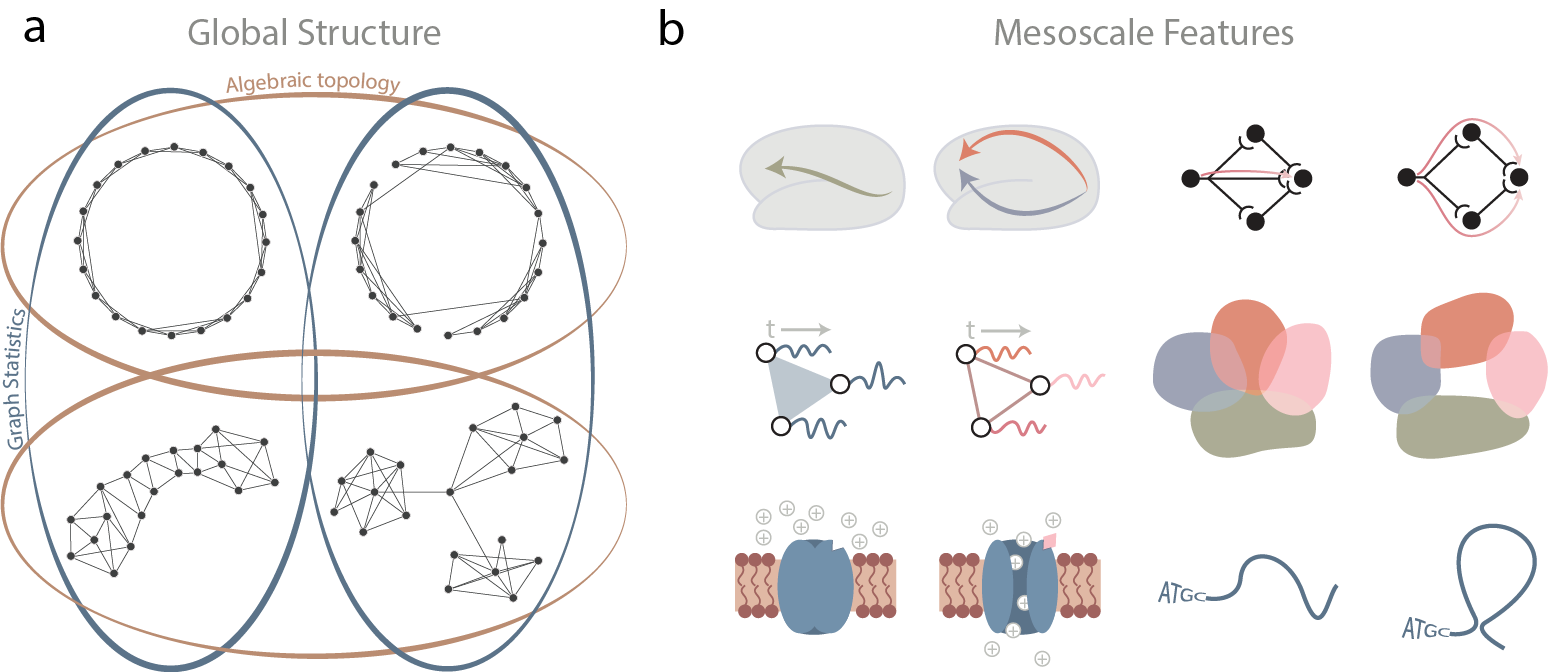}
	\caption{\textbf{Topology at the global and meso- scales.} \textbf{(a)} An illustration of four graphs that can be separated into two different sorts of groups based on either algebraic topology or graph statistics. Algebraic topology (gold) sees the circular nature of the top two graphs as distinct from the linear or star-like global structures in the bottom two graphs. In contrast, graph statistics (blue) sees the graphs on the left side as being similar, because they have the same degree distribution and no modular structure; and it also sees the graphs on the right side as being similar because they both have three modules and share the same degree distribution. \textbf{(b)} Examples of topologically trivial and non-trivial features at different scales. \emph{(Left, top)} Direct versus independent paths for information processing. \emph{(Left, middle)} Three oscillators whose dynamics are correlated in an all-to-all fashion versus in a pairwise only fashion. \emph{(Left, bottom)} A ligand-gated channel that is closed versus open to allow the flow of ions. \emph{(Right, top)} Linear versus reinforced activity flow. \emph{(Right, middle)} Neural place fields covering the entire space versus leaving gaps. \emph{(Right, bottom)} Linear DNA versus a long-range 3D interaction.}
	\label{fig0}
\end{figure}

Algebraic topology generally concerns itself with the ``shape" of topological spaces, or -- more precisely -- those properties of a space that remain invariant under stretching and shrinking (note: for a more formal treatment of the subject, the interested reader is advised to consult \cite{hatcher2002algebraic,munkres2000topology,ghrist2014elementary}). Focusing on these invariants provides a perspective that is fundamentally distinct from that offered by more geometrically-informed measures, generally termed ``network topology". As an example, Fig.~\ref{fig0} shows four graphs, which can be thought of as topological spaces; the top row contains globally circular graphs and the bottom row contains globally non-circular graphs. Algebraic topology sees (i) that the graphs in the top row are similar because they are organized into one circular loop that surrounds a ``hole", and (ii) that the graphs in the bottom row are similar because they are not organized into such a global, ``hole"-enclosing loop. 

Algebraic topology has been usefully applied at macro- and micro- scales to elucidate the function of neural systems. For example, both place fields and neural codes contain and are driven by fundamentally topological notions \cite{curto2017can,curto2008cell}. Topological data analysis has also been used with modeled spiking of hippocampal neurons generated from a simulated rat trajectory through space to suggest optimal parameters for learning the mapped space \cite{dabaghian2012topological}. Additionally \cite{arai2014effects} used topological methods to test the influence of $\theta$ band precession on learning on spatial computational models. Furthermore \cite{giusti2015clique} employed these methods to determine whether correlations of recorded pyramidal neuronal activity in the rat hippocampus contain a natural geometric structure.  At the macro scale, persistent homology has detected cyclic motifs thought to be paths of parallel information processing in the structural network \cite{sizemore2017cliques} and in networks constructed from functional data has used cycles to study learning \cite{stolz2014computational} and to distinguish drug from placebo \cite{petri2014homological}.

The consideration of topology in data analysis is relatively new \cite{carlsson2009topology}, though its methods are quite appropriate given their freedom from coordinates and robustness to noise. As with any new application of pure mathematics, mathematicians are continually pushing the technology forward and biologists are finding new questions that can now be addressed with these tools. However, since the formal mathematics underpinning topology is likely to be relatively unfamiliar to many neuroscientists, a non-trivial start-up cost may be required in order to accurately and fruitfully apply the tools to any particular neural system. Here we provide an exposition that aims to lower this activation energy by offering a basic introduction to the primary mathematics behind a main tool in topological data analysis, and examples of previous and possible applications of the tool in neuroscience. 

Specifically, in this paper we will take the reader through an introduction to the mathematics of and analyses possible with persistent homology. We will restrict the mathematical definitions to the particular case of basic persistent homology applications. To demonstrate concepts, we will perform the related analyses on three datasets covering multiple species and multiple spatial scales of inquiry: the mouse connectome \cite{rubinov2015wiring,oh2014mesoscale}, electrical and chemical synapses in \textit{C. elegans} \cite{varshney2011structural}, and genomic interaction data collected from Hi-C experiments \cite{van2010hi}. Finally we will discuss further applications and possibilities, highlighting new questions answerable with up-and-coming topological methods. We emphasize that for the interested reader, more detailed reviews and tutorials are available \cite{edelsbrunner2008persistent,ghrist2017homological}, as well as an overview of available software packages \cite{otter2017roadmap}.

\section*{When should we use topological data analysis?}

Before delving into the mathematics, it is important to first pause and ask if the system of interest is well-suited for these methods and how topology might be interpreted in this system. We suggest two main considerations when evaluating the application of topological data analysis to a specific scientific enquiry. First, are higher order interactions (more than pairwise) important in the system? Examples of such higher order interactions are evident across spatial scales in neuroscience: groups of brain regions may serve a similar function, proteins may combine into complexes, or neurons may operate in groups. For such systems, one can apply topological data analysis methods to obtain a picture of the global topology, which in turn better reveals the system's intrinsic structure \cite{giusti2015clique} and further distinguishes that structure from noise \cite{kahle2014topology,sizemore2016classification,petri2013topological,horak2009persistent}. 

Second, what does a topological cavity mean in the system under study and why might such a feature be important? Perhaps the system is the structural connectome, where a cavity in diffusion MRI could indicate axonal dropout \cite{cagatayphdthesis}. In networks constructed from functional imaging data, a cavity might arise when multiple small groups of regions display correlated activity but when the regional activity profiles are not all together similar. In chromatin, previous studies have shown that long range interactions regulate gene expression \cite{lieberman2009comprehensive,sanyal2012long} and disruption to or changes in these interactions might play a role in disease \cite{ahmadiyeh20108q24,kleinjan2005long}. We include these and a few other examples in Fig.~\ref{fig0}b, but note that creativity continues to fuel many more interpretations than can be here expressed.

\section*{Pieces and parts of the simplicial complex}

\begin{figure}[h]
	\centering
	\includegraphics[width=5in]{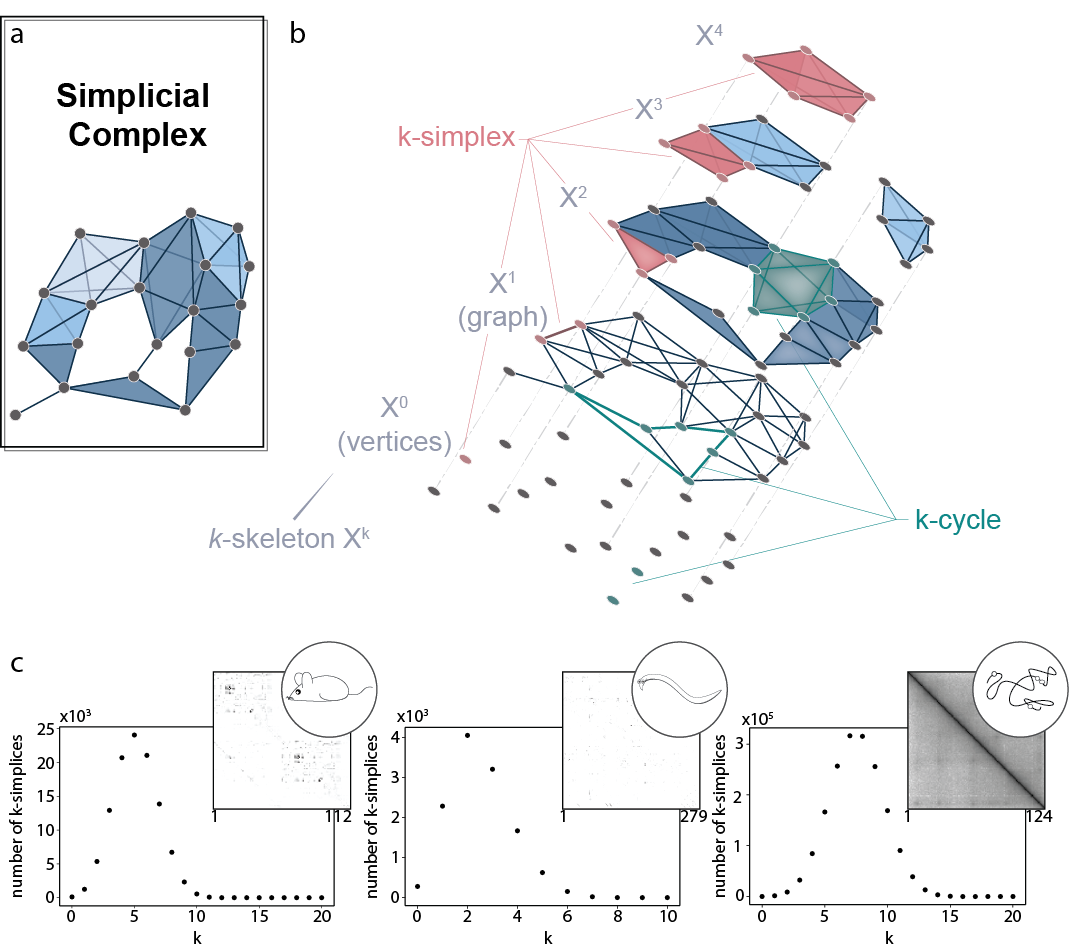}
	\caption{\textbf{The simplicial complex.} \textbf{(a)} A simplicial complex with simplicies colored by dimension. \textbf{(b)} Peering into the simplicial complex by separating it into the $k$-skeleta. An example $k$-simplex is colored in burgundy, an example $k$-chain is colored in salmon, and several example $k$-cycles are colored in teal for each dimension. \textbf{(c)} The distribution of $k$-simplices in the clique complex created from the \emph{(left)} mouse inter-areal connectome, \emph{(middle)} \textit{C. elegans} electrical and chemical synapses, and \emph{(right)} genomic interaction Hi-C data indicating distances between points on the linear genome. The adjacency matrix associated with each system is displayed in the upper right corner of each plot.}
	\label{fig1}
\end{figure}

In network science, we might commonly translate our data into a binary graph and then proceed with techniques anchored in graph theory. For topological data analysis, we instead translate our data into an object called a \textit{simplicial complex} (Fig.~\ref{fig1}a), which allows us to draw from an expanded pool of topological methods. Instead of exclusively recording the presence or absence of pairwise relations, here relational objects can connect any number of nodes. Such a unit of $k+1$ nodes is called a $k$-simplex (Fig~\ref{fig1}b, salmon), which is geometrically the convex hull of its $k+1$ affinely positioned vertices: one node is a $0$-simplex, an edge is a $1$-simplex, a filled triangle is a $2$-simplex, and so on. We write a $k$-simplex formed with nodes $v_0,\dots, v_k$ as $\{v_0,\dots, v_k\}$. A collection of nicely constructed simplices forms a simplicial complex: a set of nodes $V$ and a collection $K$ of simplices subject to the rule that if $s$ is a simplex in $K$ and $s' \subseteq s$, then $s' \in K$. This rule prevents having a 2-simplex in the complex without containing its constituent edges (1-simplices) or vertices (0-simplices), and similar atrocities. Finally it may be convenient to reference particular layers or dimensions of the simplicial complex. Formally this is the \textit{$k$-skeleton}, denoted $X^k$, or the collection of all simplices with dimension at most $k$ in the simplicial complex (Fig.~\ref{fig1}b, gray). Looped patterns of $k$-simplices called \emph{$k$-cycles} (Fig.~\ref{fig1}b, green) are of particular interest within the complex and will be more formally defined in the coming sections.

After constructing a simplicial complex (see Box 1), we can obtain a first glimpse of the system's structure by calculating the simplex distribution (Fig.~\ref{fig1}c). The number of simplices in each $k$-skeleton can provide insight into how the system functions \cite{reimann2017cliques} such as the size of functional units or can help distinguish structure from noise \cite{sizemore2016classification}. In a graph reflecting closeness of points in space, one expects that if two edges connect three nodes, the third edge closing the triangle is likely to exist. As a higher-dimensional analogue, if we construct a simplicial complex from data that has geometric constraints such as in the genomic interaction data (two nodes closer along the linear DNA are more likely to be physically near each other), we will see many larger simplices. One can even generalize many network measures, including centrality, to simplicial complexes in order to gain further insight into the system's structure \cite{estrada2017centralities}.

\begin{figure}
\begin{framed}
	\textbf{From data to simplicial complex}
	
	There are multiple ways to encode data into a simplicial complex. If the data is already in a binary graph form, the most basic representation is called the \textit{clique complex}. The clique complex (or \defn{flag complex}) is the simplicial complex formed from assigning a $k$-simplex to each $(k+1)$-clique in the graph (see panel \emph{(a)} of this box, below). This representation is generally favored when no further information beyond pairwise relations is known, since it requires the fewest choices in the designation of simplices. 
	
A second interesting way to create a complex from data can be used when higher order relations are known, such as groups of nodes sharing a feature. In this case, one can create the \defn{nerve complex} which is a \textit{$\#$simplices $\times$ $\#$vertices} matrix recording the vertices that form each simplex (corresponding to a feature) in the complex. Note that we assume downward completion of simplices; that is, if $\{v_i,v_j,v_k\}$ is a simplex then we include $\{v_i,v_j\},\{v_j,v_k\}$, $\{v_i,v_k\}$, and so on, as simplices as well. This representation is particularly suitable for data that comes in the form of nodes and groups, since in this object, one could have three edges connect in a triangle without filling in the encompassed area (i.e. we could have three 1-simplices arranged in a triangle but are not part of a $2$-simplex, see panel \emph{(b)} of this box, below). An example context in which it may be fruitful to use the nerve complex is when studying correlation among system units: here, we could imagine three vertices whose activity profiles are correlated in pairs, but not correlated all together (see Fig.~\ref{fig0}b). See \cite{giusti2016two} for details on this example and for a few additional complexes that can be used for data encoding.

If the data instead comes as a point cloud, such as points in the brain's state space, we can build a simplicial complex by choosing some value $\epsilon$ and drawing in a $k$-simplex for $k+1$ nodes that are a pairwise distance $<\epsilon$ apart. This representation is called the \defn{Vietoris-Rips} complex (see panel \emph{(c)} in this box, below) and is used to infer the shape of data \cite{carlsson2009topology}. As an extension, if points have an associated size or other feature (for example, the van der Waals radius for points representing atoms in a protein crystal structure \cite{gameiro2015topological}) one can create the \textit{alpha complex} where the parameter $\epsilon$ is weighted based on the point size \cite{edelsbrunner1995union}. 

\centering
	\includegraphics[width = 5in]{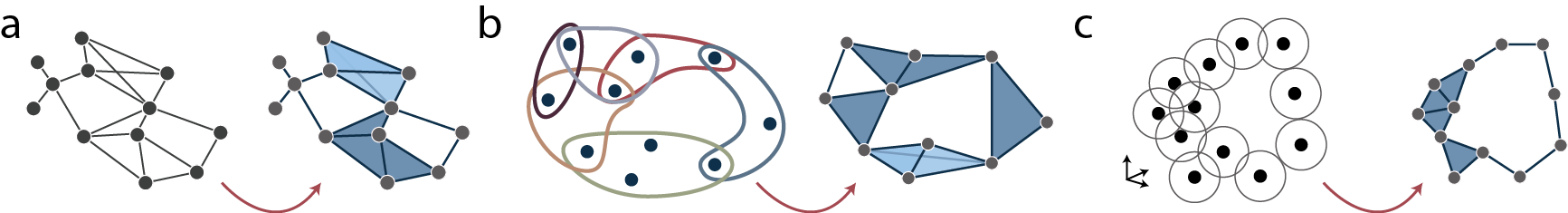}
	\caption{\textbf{Simplicial complexes from data.} \emph{Left} Clique complex created from a binary graph. \emph{Middle} Nerve complex from overlapping groups of nodes. \emph{Right} Vietoris-Rips complex from a point cloud.}

\end{framed}
\end{figure}

\section*{Topological feature compression}

We next describe how we will leverage all of this higher-order information within the simplicial complex to reveal a system's essential topological features. We begin with the simple observation that simplices contain other smaller simplices and that this containment provides a notion of relation among simplices and furthermore provides additional structural information. More specifically, a simplex in a simplicial complex is \defn{maximal} if it is not contained in any larger simplex. Further, any subset of a simplex is called a \defn{face} (Fig.~\ref{fig2}a). For example, if $\{v_i,v_j,v_k\}$ is a $3$-simplex, it contains a $2$-simplex $\{v_i,v_j\}$, where $\{v_i,v_j\}$ is a face of $\{v_i,v_j,v_k\}$. We can easily keep track of facial relations within a simplicial complex which can be used to highlight important properties, such as which $k$-simplices are maximal and which are faces of higher dimensional simplices.

\subsection*{Chain groups and boundaries}

We imagine cavities within simplicial complexes as akin to bubbles under water. Now a bubble must first and almost trivially be void of water, and secondly must be encapsulated by a water shell (the bubble's surface). Contrastingly the air completely above the water is not in a water bubble, since it does not live in a shell of water. In the same way, cavities in simplicial complexes must be void of higher simplices as well as completely enveloped, or bounded, by simplices. As we will see below, this notion of a boundary is crucial to detecting such topological features. 
 
Let's begin with a familiar matrix that records node membership within edges. This is a $\# nodes \times \# edges$ binary matrix with a 1 in position $i,j$ if the $i^{th}$ node is involved in the $j^{th}$ edge (in simplicial complex terms we would say 0-simplex $\{v_i\}$ is a face of 1-simplex $\{v_i,v_j\}$). We will call this matrix $\partial_1$. Recall that any matrix is a linear map between vector spaces. Here we are mapping from the vector space formed by assigning one basis element per edge to the vector space created from basis elements corresponding to nodes. These vector spaces are called the first and zeroth \defn{chain groups}, $C_1$ and $C_0$, respectively, and their elements (vectors) also called 1- or 0-chains. For simplicity in computations, we use vector spaces with binary coefficients so that all vectors record ``0" or ``1" in slots and vector addition proceeds via binary arithmetic: edges and vertices are either ``off" or ``on".
 
This matrix houses a large amount of structural information. First, if we send a 1-chain (vector) corresponding to one edge through this map it is sent to its two end nodes, or its boundary. Not surprisingly, $\partial_1$ is called the \defn{boundary operator}. Second, note that we can make 1-chains in $C_1$ corresponding to paths and loops (or cycles) in the simplicial complex. In a cycle, the beginning and end nodes are the same so its boundary is null. If we recall the kernel of a matrix is the subspace formed by all vectors sent to 0 by the matrix, one can show the kernel of $\partial_1$ is precisely the subspace spanned by these cycles. If the kernel of $\partial_1$ has significance, what then of the image, $\partial_1(C_1) \subseteq C_0$? In fact the image consists of all 0-chains that coincide with the nodes at the \emph{boundary} of edges and paths in the complex. 
 
\begin{figure}
	\centering
	\includegraphics[width = 5in]{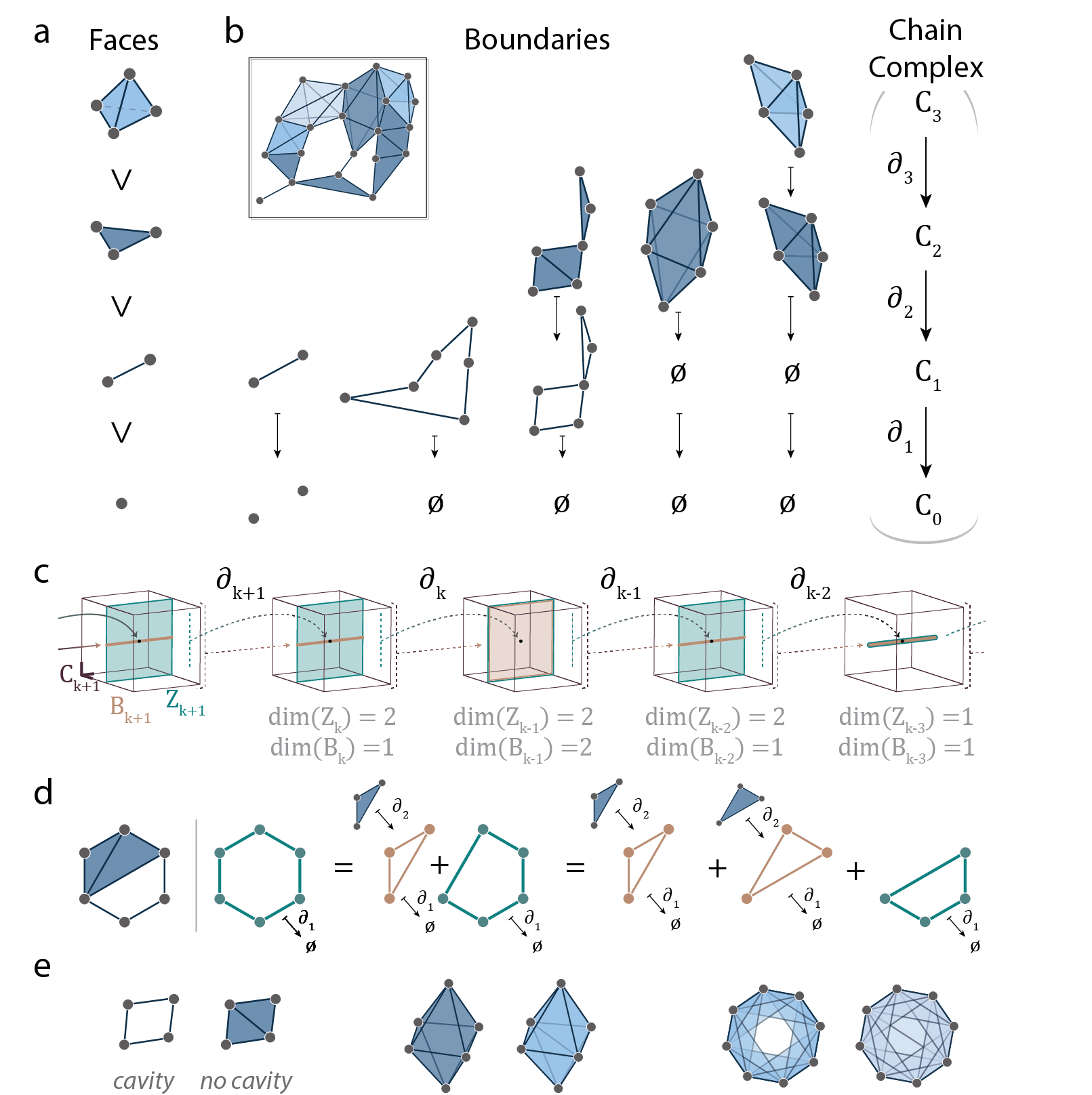}
	\caption{\textbf{Chain Complexes and Homology.} \textit{(a)} A $3$-simplex \emph{(top)} has $2$-simplices, $1$-simplices, and $0$-simplices as faces. \textit{(b)} Boundary maps take $k$-chains to their boundaries. Examples shown for dimensions $1$ through $3$. These boundary maps connect the chain groups forming the chain complex \emph{(right)}. \textit{(c)} Depiction of a chain complex with 3-dimensional chain groups (maroon boxes). Boundaries (gold) and cycles (green) are defined using the boundary operator and may span the same space, or the boundary space may be a strict subspace of the cycle space. \textit{(d)} Given the simplicial complex shown at the \emph{(left)}, the three green cycles are equivalent because they differ by boundary cycles (gold). \textit{(e)} Examples of non-trivial $k$-cycles for $k = 1$ (left), $k=2$ (middle), and $k=3$ (right). }
	\label{fig2}
	
\end{figure}

Now we again lift from networks to complexes. If we let $C_k$ (the $k^{th}$ chain group) denote the vector space with basis elements coinciding with $k$-simplices, then we can similarly create the boundary map $\partial_k$ by creating a $\dim(C_{k-1}) \times \dim(C_k)$ binary matrix with entry $i,j = 1$ only if the $i^{th}$ $(k-1)$-simplex is a face of the $j^{th}$ $k$-simplex. For example, the boundary of a $1$-chain corresponding to a path is its beginning and end vertices, the boundary of a $2$-chain corresponds to its surrounding edges, and the boundary of a $3$-chain corresponds to a shell of $2$-simplices (see Fig.~\ref{fig2}b for examples). As in the lower-dimension case the boundary operator $\partial_k$ has as a kernel $k$-cycles and sends $k$-chains to their boundaries, which must be $k-1$ cycles. Together, the chain groups $C_k$ and boundary operators $\partial_k:C_k \rightarrow C_{k-1}$ form a \defn{chain complex} (Fig.~\ref{fig2}b, right) in which the image of $\partial_k$ in $C_{k-1}$ (everything that gets hit by $\partial_k$) is contained within the kernel of $\partial_{k-1}$. We think of the chain complex as an instruction manual for assembling the simplicial complex from a collection of simple building blocks: simplices. 
 
To summarize: we can create matrices $\partial_k$ between vector spaces of $k$ and $(k-1)$-simplices that encode the structure of the simplicial complex, including its cycles and boundaries. 
 
  \[
  \boxed{\dots \xrightarrow{\partial_{k+3}} C_{k+2} \xrightarrow{\partial_{k+2}} C_{k+1} \xrightarrow{\partial_{k+1}} C_k \xrightarrow{\partial_k} C_{k-1} \xrightarrow{\partial_{k-1}} C_{k-2} \xrightarrow{\partial_{k-2}} \dots}
  \]

\subsection*{Homological algebra}

Our goal is to find the topological features -- cavities of different dimensions -- of the simplicial complex. We are close: we know how to find cycles of dimension $k$, namely $Z_k = \ker \partial_k$ or the \defn{$k$-cycle subspace}. If there are topological cavities in the simplicial complex, some of these cycles will surround them, but still other cycles will instead be boundaries of higher dimensional simplices. Those $k$-cycles that are boundaries of $(k+1))$-chains form the \defn{$k$-boundary subspace} $B_k$, a subspace of $Z_k$. Intuitively, if $Z_k$ records all the $k$-cycles, $B_k$ records which of these are ``footprints'' left by higher dimensional simplices. We could attempt to study each of the boundary maps $\partial_k:C_{k} \rightarrow C_{k-1}$ in isolation, but homological algebra tells us we will gain much more by considering the entire chain complex as a whole (a lesson relearned in persistent homology).

To appreciate this fact, consider the example chain complex shown in Fig.~\ref{fig2}c. Linking all of these matrices together, we can see how the information in one map might overlap with information provided by the subsequent map. In this example, the image of $\partial_{k+1}$ (i.e. $B_k$) is a subspace of the two-dimensional $\ker(\partial_k) = Z_k$. So $B_k\subset Z_k$ and there is one dimension of information in $Z_k$ that $B_k$ does not see. Next observe that $Z_{k-1} = B_{k-1}$. Here all the information in $Z_{k-1}$ is contained in $B_{k-1}$. Cavities arise when we have cycles that are not also boundaries, so we need lastly to understand how to extract the information descrepancy between $Z_k$ and $B_k$.

As an example, we turn to dimension $1$ in the simplicial complex shown in Fig.~\ref{fig2}d. Each green cycle surrounds the cavity within the simplicial complex, but they differ from one another by the addition of a boundary cycle (gold). This observation is indicative of an over-arching rule: when we add boundary cycles $b \in B_k$ to cycles $\ell\in Z_k$, the resulting cycle $\ell +b$ will surround the same cavity (or cavities, or no cavities) as $\ell$. Then to extract the topological cavities, we desire the \emph{equivalence classes} of cycles $\ell \in Z_k$ where two cycles $\ell_1,\ell_2 \in Z_k$ are equivalent if $\ell_1 = \ell_2 +b$ for some $b\in B_k$. Recall that the \defn{equivalence class} of an element $\sigma$ is the set of all elements equivalent to $\sigma$ and denoted $[\sigma]$. So any cycle $\ell\in [\text{green cycle}]$ will surround the cavity within this example simplicial complex.

The cycle space $Z_k$ contains a large amount information about the structure \cite{shanker2007graph} but it is more than we need in this case. Since adding boundary cycles to a given cycle does not change the cavities it surrounds, then all of the topological information we seek is stored in the cycle space that is not altered by movements along the dimensions within the boundary subspace. Formally compressing the cycle space in this way to get the equivalence classes of $k$-cycles is called taking the vector space quotient: here $Z_k/B_k =: H_k$ is called the $k$-th \defn{homology group} of the simplicial complex. That is, any boundary cycle ($b \in B_k$) acts like a 0 in the resulting space. Then the number of cavities of dimension $k$ will be the dimension of $H_k$, since $H_k$ is generated by equivalence classes of $k$-cycles, with each of these non-trivial equivalence classes corresponding to a cavity within the simplicial complex. Indeed by examining the simplicial complex shown in Fig.~\ref{fig2}c, we see that one cavity is enclosed by $1$-simplices, and by direct calculation one can verify $\dim H_1 = 1$ with the one non-trivial equivalence class represented by any one of the green cycles. 

The sequence of linear maps in a chain complex give rise to a sequence of homology groups that are the ''compressed version" of the chain complex. Turning now to Fig.~\ref{fig2}e, we see the sequence of homology groups that are the ``compressed version'' of the chain complex in Fig.~\ref{fig2}c. The dimension of the $k$-th homology group counts the number of topological cavities enclosed by $k$-cycles (including $H_0$, which counts the number of connected components). We show examples of cavities and the lack thereof in Fig.~\ref{fig2}f for dimensions 1 through 3. The dimension of the $k$-th homology group $\beta_k = \dim(H_k) = \dim(Z_k) - \dim(B_k)$ is called the $k$-th \defn{Betti number}. Betti numbers are topological invariants of the simplicial complex and are related to the well-known Euler characteristic via $\chi = \sum_{i=0}^{\infty}(-1)^i\beta_i$.

\section*{Homology from complex to complex: Persistent homology}

When considering data analysis, we have thus far discussed binary simplicial complexes such as the clique complex of a binary graph. However, biological relations often have intensities manifesting as weights on these relations (for example, streamline counts betweeen brain regions, functional similarity between neuronal activity time series, co-expression of genes, etc.). Answering the question of how to optimally (or even adequately) incorporate this additional information remains a challenge. One approach for analyses of weighted graphs is to choose a threshold on the edge weight, and to retain only edges above this threshold. However, this approach requires that one make a very strict decision about which edges are relevant and which edges are not. Here we will circumvent this choice by thresholding the weighted graph at all values to obtain a sequence of binary simplicial complexes and as a result, we will be able to follow cavities throughout the parameterized family.

\subsection*{Filtrations}

Imagine flipping through a $z$-stack or a series of two dimensional images of a cell from the top to the bottom. You might mentally note the positions of organelles, membranes, etc., in slice $i$ to encode how they map into slice $i+1$. We are going to use this idea to ``flip through'' levels of a weighted graph.

\begin{figure}
	\centering
	\includegraphics[width = 5in]{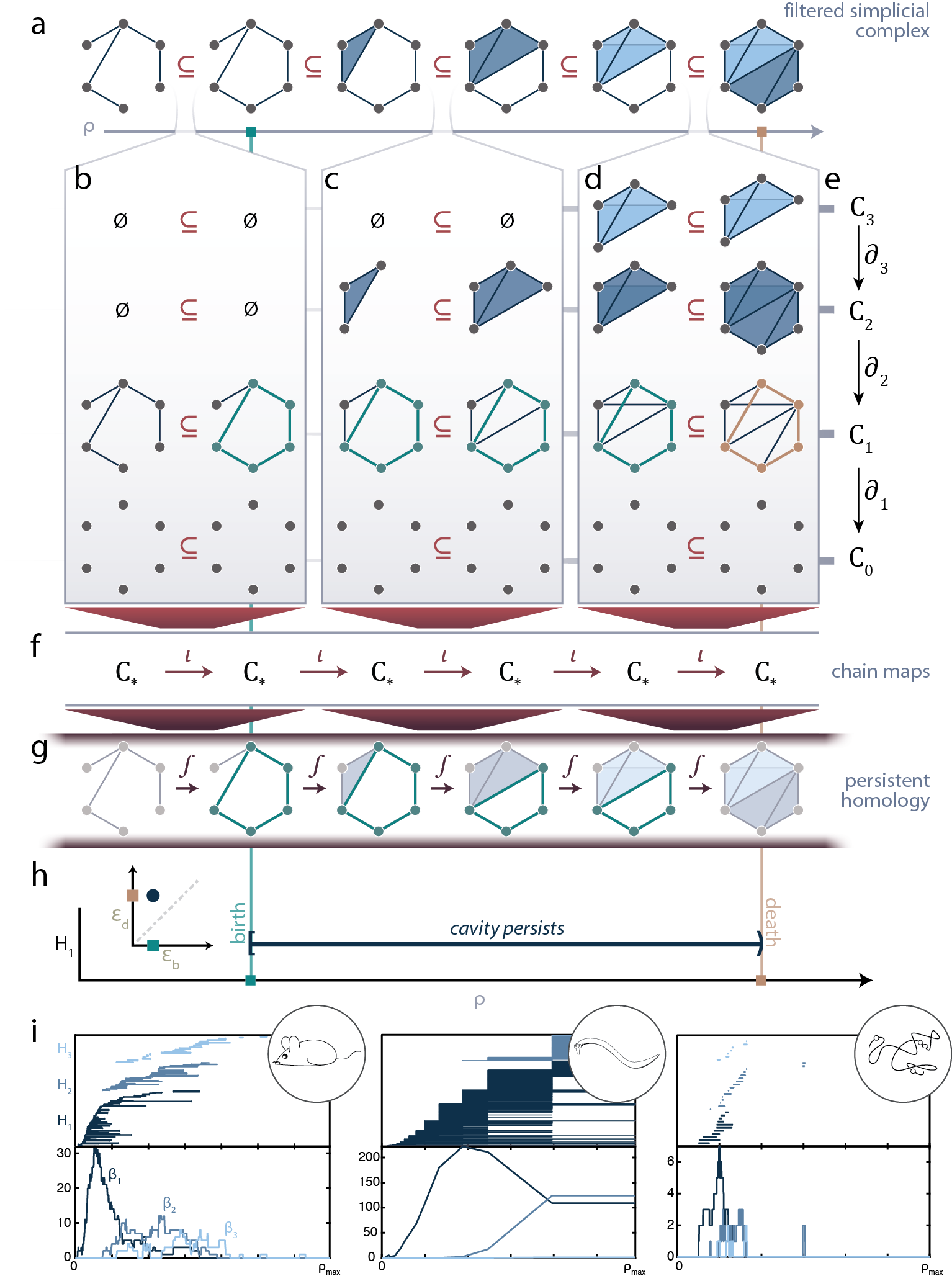}
	\caption{\textbf{Persistent Homology.} \textit{(a)} Filtered simplicial complex along the parameter $\rho$. \textit{(b,c,d)} Considering the filtration along each dimension. A persistent cycle (green) is born in panel \textit{(b)}, persists through panel \textit{(c)}, and dies in panel \textit{(d)}. Each simplicial complex has an associated chain complex \textit{(e)}. Then we get $\iota$ that map chain complexes to chain complexes \textit{(f)}. Finally these maps induce maps $f$ between homology groups \textit{(g)}. \textit{(h)} We can record the birth and death time of this example persistent cycle (green cycle as representative, panel \emph{(g)}) as a barcode or persistence diagram (inset). \textit{(h)} Barcodes \emph{(top)} and Betti curves \emph{(bottom)} for the mouse connectome \emph{(left)}, \textit{C. elegans} electrical and chemical synapses \emph{(middle)}, and genomic interaction data \emph{(right)}.}
	\label{fig3}
	
\end{figure}

In practice, we often begin with a model of a system as an edge-weighted graph, from which we can simply derive an ordering of the edges from greatest to least. We can then add edges to the empty graph following this ordering, resulting in a sequence of binary graphs where each graph in the sequence is a subgraph of the next graph in the sequence. What we have just described is an example of a \defn{filtration}, or a sequence of objects $G^0,G^1,\dots$ with each $G^i\subseteq G^{i+1}$. From each binary graph we can construct the clique complex (see Box 1) which finally induces the desired filtration of simplicial complexes.  Though creating a filtration from a weighted graph in this way is common (called the order complex \cite{giusti2015clique,sizemore2016classification} or weight rank clique filtration \cite{petri2013topological,petri2014homological}), all we really need is a weighed simplicial complex in which weights $|.|$ on each simplex follow the rule $s' \leq s \implies |s'|\leq |s|$ (for an example with weights on nodes rather than edges, see \cite{sizemore2017knowledge}). Then we can create a sequence of simplicial complexes adding simplices one at a time in order of decreasing weight, indicated by the parameter $\rho$ in Fig.~\ref{fig3}a. 

Creating a filtered simplicial complex from a weighted simplicial complex -- such as that constructed from a weighted graph -- is quite useful. It relieves the burden of finding analysis techniques that extend to weighted simplices, and it also preserves the relational information contained in the original weights by encoding such information in the ordering. This dependence on the order of weights, rather than on the numerical value of the weights, can prove particularly useful in the study of empirical recordings in which the raw measurement values are not fully reliable. It is also of course possible to map results back to the weights if originally measured empirical values are critical. 

Now we have a filtered simplicial complex with complex $K^0\subseteq \dots \subseteq K^T$, and we know how to map simplices from the $i^{th}$ complex $K^i$ into the next complex $K^{i+1}$: we send each simplex $s \in K^i$ to its natural counterpart in $K^{i+1}$ (Fig.~\ref{fig3}a). We break down this inclusion by dimension in Fig.~\ref{fig3}b-d. Recall that from each simplicial complex along the filtration, we can create a chain complex (Fig.~\ref{fig3}e). Then if we can send $k$-simplex $s$ in $K^i$ to the $k$-simplex $f(s)\in K^{i+1}$, we also immediately get maps from $C_k(K^i)$ to $C_k(K^{i+1})$ because we know how to map the basis elements: they correspond to simplices. These are called chain maps $\iota:C_k(K^i) \rightarrow C_k(K^{i+1})$ (Fig.~\ref{fig3}f) and they are defined from our knowledge of $K^i\subseteq K^{i+1}$. Thus, we can nicely map simplices to simplices, and paths to paths (see Fig.~\ref{fig3}b-d), and we can similarly map $k$-chains to $k$-chains across the filtration (Fig.~\ref{fig3}f).

\subsection*{Persistent homology}

While the mathematics that we have thus far discussed are interesting in and of themselves, the true impact of topological data analysis appears when passing from a single complex to an evolving filtration. Note in Fig.~\ref{fig3}e that there is a reminder about the chain complex with boundary maps from $C_k(K^i)\rightarrow C_{k-1}(K^i)$. Using panel \emph{(d)} as a visual example, we note that if we map an element from the 2-skeleton across to the 2-skeleton of the next complex and then down to the 1-skeleton, we get the same result as when we instead map first down to the 1-skeleton and then across to the next complex. When we move to chain maps, though less easy to directly visualize, the same property holds; that is, going across and then down is the same as going down and then across. If we think of chain complexes as pieces with assembly rules, then performing the construction and then moving to the next chain complex is the same as moving to the next chain complex and following these new, but compatible, assembly instructions.

How does this process relate to topological compression? Say that we have two equivalent 1-cycles $\ell_1 \sim \ell_2 \in C_1(K^i)$. Then we know that they must still be equivalent in $C_1(K^{i+1})$ since boundaries map to boundaries and cycles to cycles and thus we can map equivalence classes of cycles from one complex to the next. In moving from one complex to the next, we might form a new cycle (Fig.~\ref{fig3}b, green), map a non-trivial cycle to a non-trivial cycle (Fig.~\ref{fig3}c, green), or map a non-trivial cycle to a boundary cycle (Fig.~\ref{fig3}d, green to gold). From our chain maps we get induced maps on the homology groups $f_k:H_k(K^i) \rightarrow H_k(K^{i+1})$ (visualized in Fig.~\ref{fig3}g; note that we have suppressed notation and write $H_k(K^i)$ to mean $H_k(C_k(K^i))$ for simplicity). Many equivalence classes can survive these mappings, and together this is the \defn{persistent homology} of the filtered simplicial complex.

Having maps between homology groups means that we can identify the point along the filtration at which a cavity (i.e., a non-trivial equivalence class of $k$-cycles) is first formed (Fig.~\ref{fig3}b,g, green line). Then we can follow this particular cavity as we add more simplices (Fig.~\ref{fig3}c,g), and finally we can know the point at which this cavity is filled (Fig~\ref{fig3}d,g, the non-trivial equivalence class maps to the trivial equivalence class of boundary cycles). It is important to note that we highlight in our illustration only one representative $k$-cycle in each equivalence class to represent the persistent cavity, though we could have chosen any equivalent cycle as a representative. The first appearance of a persistent cavity is called its \defn{birth}, the value at which it is filled in is known as its \defn{death}, and the difference between the death and birth is called the \defn{lifetime} of the persistent cavity. This process assigns a half-open interval $[b,d)$ to the persistent cavity with $b,d$ the values of the persistent cavity birth and death, respectively. Then all persistent cavities within a filtered simplicial complex can be visualized as part of a barcode (Fig~\ref{fig3}h) or a point on the extended half-plane (peristent cavities that never die are given a death value of $\infty$) called the persistence diagram of dimension $k$ (Fig.~\ref{fig3}h, inset). 

The barcode in each dimension defines a signature of the filtered simplicial complex that describes how topological features evolve along the filtration, providing insight into local-to-global organization. Importantly, a barcode gives more than a summary statistic; it shows not only the existence or number of persistent features, but also when they arise and overlap along the filtered simplicial complex. For example, the filtered clique complex of a weighted ring graph would produce one long-lived $1$-cycle, arising from the one circular loop. Since a long-persisting cavity avoids death for an extended time, we assume that long-lived persistent cavities describe more fundamental features of the complex, while those short-lived cavities are more likely due to noise in the system. Previous work has used these persistence intervals to study and classify weighted networks \cite{horak2009persistent,petri2013topological,sizemore2016classification} and to identify topological changes in cerebral arteries that track with a participant's age \cite{bendich2016persistent}.

We have computed the persistent homology of the mouse connectome, the \textit{C. elegans} electrical and chemical synapse circuit, and genomic Hi-C data (Fig.~\ref{fig3}i). We plot the barcode (above) and Betti Curves $\beta_k(\rho) = k$th Betti number at filtration parameter $\rho$. In Fig.~\ref{fig3}i we see that the mouse interareal connectome has a few long-lived $1$-cavities and little higher dimensional persistent homology, similar to the persistent homology previously reported for the structural connectome of humans \cite{sizemore2017cliques}. In comparison, the persistent homology from the \textit{C. elegans} electrical and chemical synapses shows many more persistent cavities in addition to those of higher dimensions. Finally the persistent homology of the genomic interaction data reveals only a few persistent cavities, which is expected due to the high numbers and degrees of simplices (Fig.~\ref{fig1}c). Though all three datasets are embedded in $\mathbb{R}^3$, we speculate that higher dimensional features may still play an important role in the system's function, such as in the computational capacity of the two neuronal networks.

Once the persistent homology has been computed, one can compare barcodes (or persistence diagrams) using a few different methods. First, we note that these objects are well studied even from a theoretical viewpoint for the random weighted graph \cite{kahle2009topology} and the random geometric complex \cite{kahle2011random,bobrowski2015maximally}. Thus, by visually comparing the persistence diagrams of the empirically observed structure to that expected in these well-studied models, one can deduce that the system may be totally random or geometric \cite{giusti2015clique}. Others have compared the persistence diagrams of weighted network models using cavity lifetimes, weighted integrals of Betti curves, birth times, and more \cite{petri2013topological,horak2009persistent,sizemore2016classification}. One can also compute the \textit{bottleneck distance} between persistence diagrams \cite{cohen2007stability} or the distance between a slightly modified version of the barcode called a persistence landscape, which has nicer statistical properties \cite{stolz2017persistent,bubenik2015statistical,berry2018functional}.

\subsection*{Extracting topological features}

Since each bar in the barcode arises from a particular persistent topological feature, one might hope to extract information about the ``most important'' or ``most robust'' topological features observed. This goal is in fact a more complicated problem than might be obvious at first glance, but nevertheless a solvable and tunable one.

\begin{figure}[H]
	\centering
	\includegraphics[width = 5in]{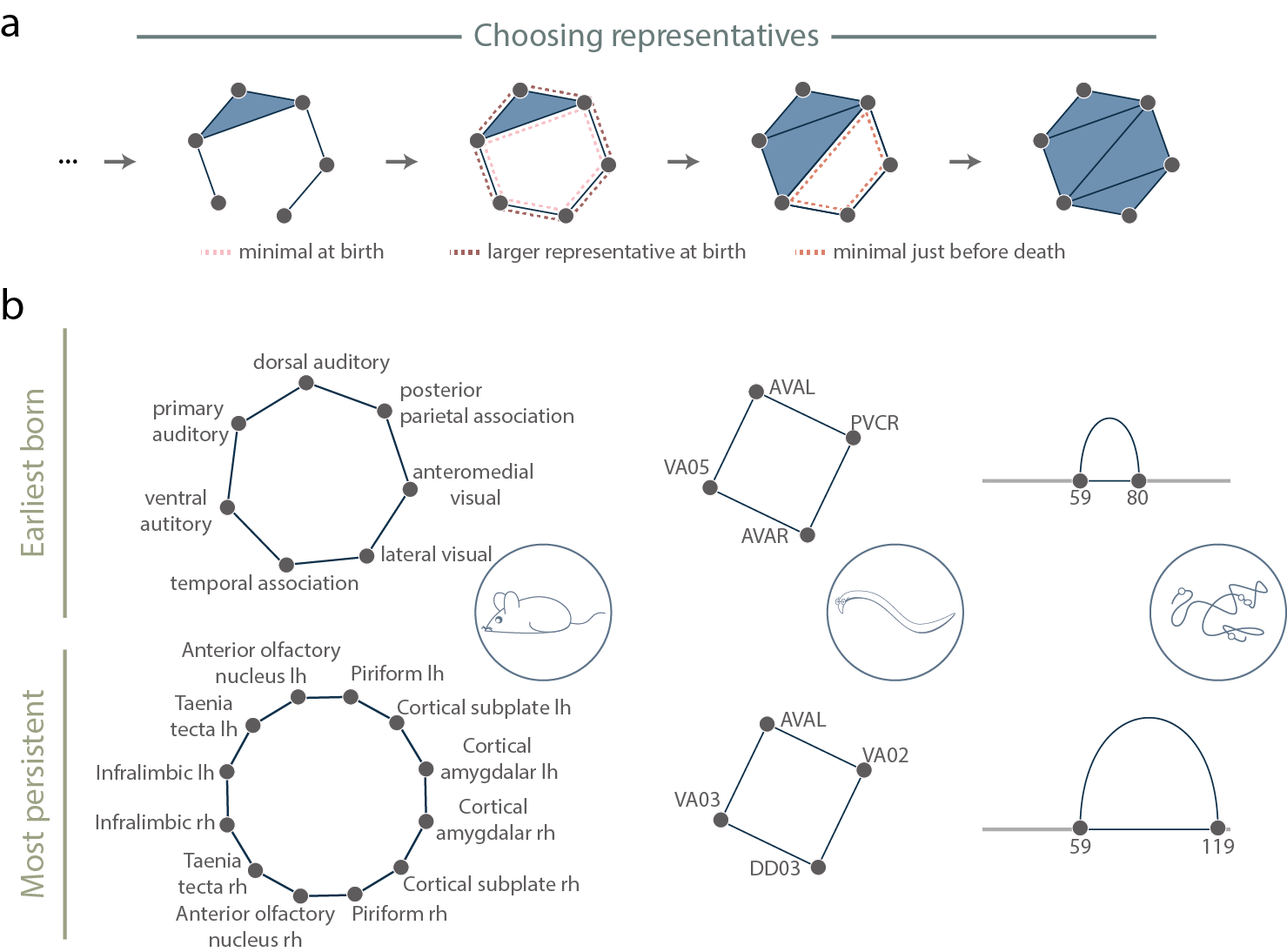}
	\caption{\textbf{Extracting generators.} \emph{(a)} Filtered simplicial complex with one persistent homology class of dimension 1. Three possible representatives are the minimal at birth (pink dashed), larger cycle at birth (brown dashed), and minimal mid-life or just before death (orange dashed). \emph{(b)} \emph{(Top)} A representative at the birth time of the earliest born and \emph{(bottom)} longest-lived persistent cavities in the mouse interareal connectome \emph{(left)}, \textit{C. elegans} electrical and chemical synapses \emph{(middle)}, and genomic Hi-C interactions \emph{(right)} with the gray line representing linear genomic position, and with the blue loop representing the nodes and edges involved in the representative cycle. }
	\label{fig4}
	
\end{figure}

First, if we only wish to determine the nodes and simplicies that surround a cavity -- how difficult could that be? Recall that in persistent homology we are mapping \emph{equivalence classes} corresponding to cavities through the filtration. This means that to extract the loop, we must \emph{choose} a particular representative (or set of representatives) from the equivalence class of a cavity-surrounding loop. Often times, the representative(s) with minimal hop distance at the birth index is(are) used (see Fig.~\ref{fig4}a, pink dashed line). We emphasize that there may exist multiple representatives with minimal hop distance, and thus care must be taken if the later analyses and interpretations require only one representative cycle per persistent homology class. For reasons related to the system under study, one may be most interested in minimal cycles at just before the death time (Fig.~\ref{fig4}a, orange dashed line), or some large cycle in the middle of the persistent cavity lifetime. Many persistent homology software implementations will also report a representative cycle for each persistent homology class, though often this is used for computing the barcodes only and there are no guarantees that the cycle that is identified is geometrically nice \cite{henselmanghrist16,adams2011javaplex}. However, methods now exist to find the generator optimizing some other parameter, for example a given weighting on the simplices \cite{dey2011optimal}, persistence \cite{busaryev2010tracking}, or minimal length along the genome \cite{emmett2016multiscale}. No one method may be ideal for all analyses, and instead the choice of method represents an area of optimization tunable to the system under study. Future work may expand this set of methods through the development of additional novel algorithms.

The second part of this question is how to determine if a persistent cycle is significant. As noted above, we generally consider the longest-lived persistent cycles to be the most intrinsic to the filtered simplicial complex. Additionally, we may find cavities appearing much earlier in the filtration than expected given some null model \cite{bobrowski2015maximally,kahle2013limit,sizemore2017cliques,stolz2017persistent}. To illustrate these ideas, we show in Fig.~\ref{fig4}b a representative cycle at birth from the earliest born \emph{(top)} and longest-lived \emph{(bottom)} persistent cavities in our datasets. For the sake of the pedagogical nature of this tutorial, we keep our calculations simple and report cycles returned by the Eirene software \cite{henselmanghrist16}. For the structural brain network, one might argue that the earliest born cavities are the most crucial since they involve the highest-weighted edges of the system. For genomic interaction data, in contrast, at this particular scale the most persistent cycle could be more interesting since it can indicate larger long-distance interactions \cite{emmett2016multiscale}.

\section*{Conclusion}

The language of algebraic topology offers powerful tools for network scientists. Using three example datasets we illustrated how to translate a network into a simplicial complex and both apply and understand persistent homology, concluding with results and reasonable interpretations. Now, for the interested network neuroscientist, we revisit our initially posed questions, hoping that our deeper understanding of the subject will spur the generation of new ideas. What does a cavity mean in any given scientist's system of interest? How would one define small groups making functional units? As with any method it is important to understand both the fundamental assumptions and underlying mathematics.

Moreover, we can push even further than the methods demonstrated in this paper. Applied topology is only about two decades old, but has nevertheless captured the attention of scientists and mathematicians alike. We have since charged past the initial applications and can now use topology to answer more complicated and detailed questions. Does the system of interest have weighted nodes instead of weighted edges \cite{sizemore2017knowledge}? Is the system time-varying with either unweighted \cite{botnan2016algebraic,carlsson2010zigzag} or weighted edges \cite{cohen2006vines,munch2013applications,yoo2016topological}? Would one perhaps wish to obtain circular coordinates for the data \cite{de2011persistent,rybakken2017decoding}? Would equivalence classes of paths instead of cycles be useful \cite{chowdhury2018persistent}? Does one need distances between networks \cite{chowdhury2015metric} or multiple filtering parameters \cite{carlsson2009computing,lesnick2015interactive}? These and more questions are recently answerable using concepts developed in applied topology. 

To conclude, the translation of data into the language of algebraic topology opens many doors for analysis and subsequent insight. Those scientists with an understanding of this language can add a myriad of powerful tools to their analysis arsenal. Additionally, continued discussion between the mathematicians developing the tools and the scientists applying the tools will continue to spur methodical advances with biological questions as the driving force. We hope from these local interdisciplinary collaborations that a larger, global understanding of neural systems will emerge.

\section*{Acknowledgments}
The authors would like to thank Richard Betzel, Harvey Huang, and Sunnia Chen for helpful discussions. D.S.B. and A.E.S. acknowledge support from the John D. and Catherine T. MacArthur Foundation, the Alfred P. Sloan Foundation, the ISI Foundation, the Paul Allen Foundation, the Army Research Laboratory (W911NF-10-2-0022), the Army Research Office (Bassett-W911NF-14-1-0679, Grafton-W911NF-16-1-0474, DCIST- W911NF-17-2-0181), the Office of Naval Research, the National Institute of Mental Health (2-R01-DC-009209-11, R01 – MH112847, R01-MH107235, R21-M MH-106799), the National Institute of Child Health and Human Development (1R01HD086888-01), National Institute of Neurological Disorders and Stroke (R01 NS099348), and the National Science Foundation (BCS-1441502, BCS-1430087, NSF PHY-1554488 and BCS-1631550). The content is solely the responsibility of the authors and does not necessarily represent the official views of any of the funding agencies.

\section*{Author Contributions}
A.E.S and D.S.B designed and directed the project. A.E.S. performed the analysis and drafted the manuscript. J.P.C. supplied Hi-C data. All authors contributed to the final version of the manuscript.


\newpage
\bibliographystyle{plain}
\bibliography{bibfile}


\end{document}